\documentclass[preprint2]{aastex}

\shorttitle{NEBULAR spectral synthesis code}
\shortauthors{Schirmer}

\usepackage{amssymb}
\usepackage{amsmath}
\usepackage{natbib}
\usepackage{rotating}
\usepackage{pdflscape}

\newcommand{\ergscmdHz}{erg\,s$^{-1}$\,cm$^{3}$\,Hz$^{-1}$}
\DeclareRobustCommand{\ion}[2]{\textup{#1\,\textsc{\lowercase{#2}}}}

\begin{document}
\bibliographystyle{aa}

\title{NEBULAR: A simple synthesis code for the hydrogen and helium nebular spectrum}


\author{Mischa Schirmer}
\affil{Gemini Observatory, Casilla 603, La Serena, Chile}
\email{mschirme@gemini.edu}



\begin{abstract}
{\tt NEBULAR} is a lightweight code to synthesize the spectrum of an ideal, mixed hydrogen 
and helium gas in ionization equilibrium, over a useful range of densities, temperatures
and wavelengths. Free-free, free-bound and two-photon continua are included 
as well as parts of the \ion{H}{I}, \ion{He}{I} and \ion{He}{II} line series.
{\tt NEBULAR} interpolates over publicly available data tables; it can be used to
easily extract information from these tables without prior knowledge about
their data structure. The resulting spectra can be used to e.g. determine equivalent 
line widths, constrain the contribution of the nebular continuum to a bandpass, and for 
educational purposes. {\tt NEBULAR} can resample the spectrum on a user-defined wavelength 
grid for direct comparison with an observed spectrum; however, it can not be used to fit 
an observed spectrum.
\end{abstract}



\section{Introduction}
Astrophysical nebulae have complex spectra. Photoionization codes, such as {\tt CLOUDY} 
\citep{fph13}, calculate realistic spectra of nebulae for a wide range of physical parameters
of the gas and the ionizing sources near or within them. These codes may have complex setups 
and significant execution times because of their ambitious scope.

{\tt NEBULAR} is not a contender in this area, because it does not fit an observed 
spectrum, nor does it return physical parameters. However, it synthesizes 
an accurate and fairly complete spectrum of an ideal, mixed hydrogen and helium gas in 
ionization equilibrium, over a wide range of temperatures, densities and wavelengths. 
Hydrogen and helium are the most abundant elements in the Universe, and thus form a central
part in the spectral analysis of gaseous nebulae. {\tt NEBULAR} provides easy and quick
access to the most recent atomic computations of these elements. Because it is lightweight, 
it could be used in larger analysis packages. It is expedient for physical purposes and 
astronomy alike, whenever the general properties of a spectrum are investigated. For example, 
it can take the arbitrary wavelength grid of an observed real spectrum as input and compute 
a model spectrum on the same grid for direct comparison. Because of the simple installation 
and usage, it can also be used for teaching purposes.

{\tt NEBULAR} is open source, licensed under the GNU Public License (GPL) and available
on {\tt Github}\footnote{{\tt https://github.com/mischaschirmer/nebular/releases}};
this paper refers to version 1.2. {\tt NEBULAR} is written in {\tt C++} and links against 
common standard libraries, facilitating cross-platform usability and installation.

This paper is organized as follows. Methods and relevant physical relations are
recapped in Section 2, together with explanations of the code's limitations.
Section 3 gives an overview of software verification, usage, the output format,
and a few applications. A summary is given in Section 4. Three example runs
are presented in the Appendix, together with further ancillary tables. All wavelengths 
in this paper are vacuum wavelengths, and all logarithms refer to base 10.

\section{Methods and limitations of the code}
The following is a summary of the physical processes involved.

The combined hydrogen and helium spectrum is synthesized from 
four different components: the free-bound, free-free and two-photon continua,
and part of the \ion{H}{I}, \ion{He}{I} and \ion{He}{II} emission line series. 
The respective rescaled emission coefficients $j$ are written as
\begin{equation}
  j_{\rm fb}(\nu)=\frac{1}{4\pi} N_X\,N_e\,\gamma_{\rm fb}(\nu)
\end{equation}
\begin{equation}
  j_{\rm ff}(\nu)=\frac{1}{4\pi} N_X\,N_e\,\gamma_{\rm ff}(\nu)
\end{equation}
\begin{equation}
  j_{\rm 2q}(\nu)=\frac{1}{4\pi} N_X\,N_e\,\gamma_{\rm 2q}(\nu)
\end{equation}
\begin{equation}
  j_{\rm nn'}=\frac{1}{4\pi} N_X\,N_e\,\gamma_{\rm nn'}\;.
\end{equation}
Here, $N_X$ and $N_e$ are the ion and electron number densities in cm$^{-3}$,
respectively, and $\gamma$ is the emission coefficient (in \ergscmdHz; apart from
$j_{\rm nn'}$, which has no frequency dependence).

The sources of the input data, as well as the limitations of the code, are 
reviewed below.

\begin{table*}
\caption{Validity matrix for {\tt NEBULAR}. Extrapolation may be used
  outside these ranges.\label{validityrange}}
\begin{tabular}{lrrrrrrrrrrr}
\tableline\tableline
Component & Temperatures & Densities & Wavelength range \\
          & $T$ [1000 K] & ${\rm log}\,(N_e/{\rm cm}^{-3})$ & \\
\tableline
Free-bound (\ion{H}{I})   & $0.1-100$    & $-$ & $911.8$\,\AA$-36.5$\,$\mu$m\\
Free-bound (\ion{He}{I})  & $0.1-100$    & $-$ & $504.3$\,\AA$-36.5$\,$\mu$m\\
Free-bound (\ion{He}{II}) & $0.1-100$    & $-$ & $228.0$\,\AA$-36.5$\,$\mu$m\\
Free-free (all)           & $0.003-10^{7}$   & $-$ & Radio to $\gamma$-rays\\
Two-photon (\ion{H}{I})   & $0.3-100$   & $\lesssim8$ & $\geq 1215.7$\,\AA\\
Two-photon (\ion{He}{I})  & $5-20$  & $-$ & $\geq 625$\,\AA\\
Two-photon (\ion{He}{II}) & $1-100$   & $\lesssim8$ & $\geq 303.9$\,\AA\\
\ion{H}{I}   lines      & $0.5-30$  & $2-8$ & $911.7$\,\AA$-27.8$\,$\mu$m\\
\ion{He}{I}  lines      & $5-25$ & $1-14$ & $2945$\,\AA$-2.16$\,$\mu$m\\
\ion{He}{II} lines      & $0.5-100$ & $2-8$ & $227.9$\,\AA$-34.6$\,$\mu$m\\

\tableline
\end{tabular}
\end{table*}

\subsection{Free-bound continuum}\label{freebound}
Free-bound emission is caused by ions capturing free electrons. 
\citet{ers06} have tabulated the continuous free-bound recombination 
emission coefficients, $\gamma_{\rm bf}$, of 
\ion{H}{I}, \ion{He}{I} and \ion{He}{II} (the ionic states after 
recombination) over temperatures $10^2\leq T\leq10^5$\,K with a step size of
$\Delta\,{\rm log}_{10}T=0.1$ (hereafter, the logarithmic bases are omitted). 
{\tt NEBULAR} performs a cubic spline 
interpolation over temperatures, and a linear interpolation over frequencies within discontinuities. As stated in \citet{ers06},
bilinear interpolation reproduces the true values to within
1\% or better.

\subsection{Free-free continuum}\label{freefree}
Free-free emission is caused by electrons scattering 
in the Coulomb field of ions with nuclear charge $Z$. {\tt NEBULAR}
computes the free-free spectrum for H$^+$, He$^+$ and He$^{++}$.
The emission coefficients are
\begin{multline}
  j_{\rm ff}(\nu) = \frac{1}{4\pi}\,N_X\,N_e\; \frac{32\,Z^2 e^4 h}{3 m_e^2 c^3}
  \left(\frac{\pi h \nu_0}{3k_BT}\right)^{1/2}\;\times\\
       {\rm exp}(-h \nu/k_B T)\;g_{\rm ff}(Z,T,\nu)
\end{multline}
\citep[e.g.][]{brm70}.
Here, $e$, $m_e$, $h$, and $k_B$ are the electron charge, electron mass, 
Planck constant and Boltzmann constant, respectively. $h \nu_0$ denotes the
ionization energy for hydrogen. Note that this equation must be evaluated in
{\tt cgs} units, in particular the electronic
charge is $1e=4.803207\times10^{-10}$\,statC, with 
$1\,{\rm statC}=1$\,cm$^{3/2}$\,g$^{1/2}$\,s$^{-1}$.

$g_{\rm ff}(Z,T,\nu)$ is the Gaunt factor, a quantum mechanical correction to
the classical scattering formalism. For typical conditions in astrophysical 
nebulae, $g_{\rm ff}\sim1-2$. Non-relativistic Gaunt factors have been calculated by 
\citet{vhw14} for a wide range of temperatures, frequencies and nuclear 
charges. {\tt NEBULAR} uses the tabulated temperature-averaged  
$\langle g_{\rm ff}(\gamma^2,u)\rangle$ for Maxwell-Boltzmann electron 
distributions, where $\gamma^2=Z^2\,h\nu_0/(k_B T)$ and $u=h\nu/(k_B T)$.

The data for $\langle g_{\rm ff}(\gamma^2,u)\rangle$ are available on a log-log grid 
with regular step sizes in both dimensions. {\tt NEBULAR} interpolates this grid
using a 2D bicubic interpolation from the GNU Scientific Library 
({\tt GSL} v2.1).

\subsection{Two-photon continuum}\label{twophoton}
\subsubsection{\label{hhe2q}\ion{H}{I} and \ion{He}{II}}
The 2$^2$S level in hydrogenic ions can e.g. be populated by direct recombination 
onto that level, and by downward cascades following electron capture to a higher 
level. A radiative decay to the 1$^2$S level by emission of a single photon is 
forbidden by the dipole selection rules \citep{lam25}. However, the transition 
may occur if \textit{two} photons are emitted, whose energies add up to the 
Ly$\alpha$ energy. The transition probability for the hydrogenic 
$2^2{\rm S}\rightarrow1^2{\rm S}$ two-photon decay is 
\begin{equation}
A_{2q}=8.2249\,Z^6\,\frac{R_Z}{R_H}\, {\rm s}^{-1}
\end{equation}
\citep{nus84}. Here, $R_H$ and $R_Z$ are the Rydberg constants for hydrogen and for
hydrogenic ions with nuclear charge $Z$, respectively. Two-photon decays from higher 
levels \citep{hir08} are ignored, as they contribute on the sub-percent level, only.

The two-photon spectrum has a natural upper cut-off at the Ly$\alpha$ 
frequency. It peaks at half the Ly$\alpha$ frequency when expressed in 
frequency units, emphasizing the importance of this process for restframe 
near-UV and far-UV observations. For low densities, the hydrogen two-photon 
emission dominates the hydrogen free-bound continuum (Sect. \ref{freebound}) 
below 2000\,\AA\;. 

The two-photon emission coefficient is
\begin{equation}
\gamma_{2q}(\nu)=\alpha^{\rm eff}_{2^2{\rm S}}(T,N_e)\,g(\nu)\,
\left(1+\frac{\sum_X N_X\,q^X}{A_{2q}}\right)^{-1}
\end{equation}
Here, $\alpha^{\rm eff}_{2^2{\rm S}}$ is the effective recombination coefficient, $g(\nu)$ 
the frequency dependence, and $q^X$ the collision rate transition coefficient; see 
also eq. (4.29) in \citet{osf06}, and eq. (9) in \citet{sts14}. 

The three factors are discussed in the following.

\subsubsubsection{Effective recombination coefficient $\alpha^{\rm eff}_{2^2{\rm S}}$}
$\alpha^{\rm eff}_{2^2{\rm S}}(T,N_e)$ is the effective recombination coefficient to 
the $2^2$S level. It has been tabulated for both \ion{H}{I} and \ion{He}{II}
by \citet{hus87} over wide temperature and density ranges (Case B, only).
The coefficients from the original publication were extracted and digitized,
and are reproduced in the Appendix (Tables \ref{recombcoeff_hydrogen}
and \ref{recombcoeff_helium}). For convenience, these tables contain the
coefficients for the $2^2$P level as well, so that the total $n=2$ effective
recombination coefficient may be calculated (not needed for the two-photon 
emission). For \ion{H}{I}, {\tt NEBULAR} extrapolates $\alpha^{\rm eff}_{2^2{\rm S}}(T,N_e)$ 
from ${\rm log}\,T=3.0$ to 2.5, from ${\rm log}\,T=4.5$ to 5.0, and from 
${\rm log}\,N_e=2.0$ to 0.0. For \ion{He}{II}, extrapolation is used from 
${\rm log}\,T=3.5$ to 3.0, and from ${\rm log}\,N_e=2.0$ to 0.0. These extrapolations 
are expected to be accurate within a few percent, as the functional dependence 
of $\alpha^{\rm eff}_{2^2{\rm S}}(T,N_e)$ at these densities and temperatures is weak.

\subsubsubsection{Frequency dependence $g(\nu)$}
The frequency dependence of the two-photon spectrum is absorbed in
\begin{equation}
  g(\nu) = h\,\frac{\nu}{\nu_{12}}\,A(\frac{\nu}{\nu_{12}}) / A_{2q}\,.
\end{equation}
Here, $\nu_{12}$ is the Ly$\alpha$ frequency of the respective hydrogenic
ion.
\citet{nus84} have presented an analytic fit for the frequency dependent 
transition probability, $A(\nu/\nu_{12})$. It is accurate to within 1\% 
over nearly the full spectral range,
\begin{equation}
  A(\nu/\nu_{12}) = C \left[ y (1-(4y)^\gamma) + 
    \delta y^\beta\,(4y)^\gamma\right]\,,
\end{equation}
with $y=\frac{\nu}{\nu_{12}}(1-\frac{\nu}{\nu_{12}})$, 
$\beta=1.53$, $\gamma=0.8$, $\delta=0.88$ and $C=202.0\,Z^6\,R_Z/R_H$\,s$^{-1}$, 
where $Z$ is the nuclear charge. 

\subsubsubsection{Collisional de-population of the $2^2{\rm S}$ level}
The 2$^2$S level can be de-populated by means of angular momentum changing 
($l$-changing) collisions, $2^2{\rm S}\rightarrow2^2{\rm P}$, inflicted by 
other ions and electrons. The $2^2$P level may then decay to the 
$1^2$S level by emitting a Ly$\alpha$ photon, reducing the two-photon flux. This 
is described by the collisional transition rate coefficients, $q^{X}(T,N_e)$.
Rates for $l$-changing collisions at a fixed principal quantum number $n=2$ are 
much higher than for $n$-changing collisions, and $n=2\rightarrow3$ collisions 
typically require $N_e\gtrsim10^6$\,cm$^{-3}$ \citep[e.g.][]{hus87}. Therefore, 
$n$-changing collisions are ignored.

\citet{pes64} have developed a theory to calculate these collisional transition 
rates. \citet{vos12} have argued that these rates are significantly overestimated.
This, however, has been disputed by \citet{sts15} and \citet{gbw16}. Consequently, 
{\tt NEBULAR} uses the prescription of \citet{pes64}. Accordingly, the hydrogen 
two-photon emissivity becomes significantly reduced for $N_X\gtrsim1000$\,cm$^{-3}$.

\subsubsection{The \ion{He}{I} two-photon continuum}
The $1{\rm s}2{\rm s}\,^1$S and $1{\rm s}2{\rm s}\,^3$S states in \ion{He}{I} may 
also decay by two-photon emission to the ground state \citep{dvd69}. The decay 
rates for $1{\rm s}2{\rm s}\,^3$S are ignored, as they are about 10 orders of 
magnitude lower than for the $^1$S state. The total transition probability is
\begin{equation}
A_{2q}=53.1\,{\rm s}^{-1}
\end{equation}
for $1{\rm s}2{\rm s}\,^1$S. In analogy to \cite{nus84}, the spectral dependence 
tabulated in \citet{dvd69} is approximated in {\tt NEBULAR},
\begin{equation}
A(y) = C [1 + \beta(y-1/2)^2 + \gamma(y-1/2)^4],
\end{equation}
with $y=\nu/\nu_{12}$, $C=145.29$\,s$^{-1}$, $\beta=-2.259$ and $\gamma=-8.77$.
This fit is accurate to 0.9\% or better over $0.025\leq y\leq0.975$. 

The rest of the treatment is the same as outlined in Sect. \ref{hhe2q},
with the notable distinction that no extensive data are available for 
$\alpha^{\rm eff}(T,N_e)$. While {\tt CLOUDY} calculates these parameters internally, 
{\tt NEBULAR} relies on the single temperature dependence between $5000-20000$\,K 
listed in \citet{gsf97}. Fortunately, this is not a big drawback: for typical 
abundances and depending on $T$, the \ion{He}{I} 
two-photon continuum is $2-6$ orders of magnitude below the \ion{H}{I} 
two-photon continuum. Peak emission in $f_\lambda$ occurs at around 770\,nm.

\subsection{Emission lines}
\subsubsection{\ion{H}{I} emission lines}
The \ion{H}{I} emission line spectrum for Case B (Lyman-thick nebulae) is built from 
\citet{sts15} and includes all line transitions from upper levels $n\leq99$ onto lower 
levels $n'\leq8$. The latter was chosen in {\tt NEBULAR} so that the $\alpha$ line of 
a $j_{nn'}$ series 
has a wavelength shorter than 36\,$\mu$m, i.e. it still falls within the free-bound 
continuum calculated by \citet{ers06}. The parameter space covers ${\rm log}\,N_e=2-6$ 
and ${\rm log}\,T=2.0-4.4$. The Lyman series ($n'=1$) is excluded for Case B.

Linear extrapolation is used from ${\rm log}\,N_e=2$ to 0, and has been verified by 
comparison with \citet{pen64} for ${\rm log}\,N_e=0$ and $T=1250-80000$\,K. It is
accurate within 2.5\% for $T=2500-40000$\,K. For $T=1250$\,K and $T=80000$\,K, the 
errors are 4\% and 7\%, respectively. Calculations of the \ion{H}{I} line spectrum 
are limited to upper levels of $n\leq 25$ \citep{sth95} if extrapolation is used, and 
are skipped for $T>50000$\,K. 

For Case A (Lyman-thin nebulae), the emission coefficients are taken from \citet{sth95}.
The same limitations hold as outlined above; the Lyman series ($n'=1$) is included .

\subsubsection{\ion{He}{II} emission lines}
The \ion{He}{II} spectrum for Case A and Case B is based on \citet{sth95}, with a common upper
level of $n\leq25$. Treatment is the same as outlined for \ion{H}{I}, with some differences: 
First, transitions onto lower levels $n'\leq14$ are included. Second, the temperature 
range is limited to $500\leq T\leq10^5$\,K. Third, calculations are skipped if $T>10^5$\,K.

\subsubsection{\ion{He}{I} emission lines}
The data from \citet{pfs13} are used \citep[corrigendum to][]{pfs12} to construct the 
\ion{He}{I} spectrum for $T=5000-25000$\,K, ${\rm log}\,N_e=1-14$, 
and Case B. In total, 44 singlet and triplet lines are available between $2946-21623$\,\AA.

\begin{figure}[t]
\includegraphics[width=1.0\hsize]{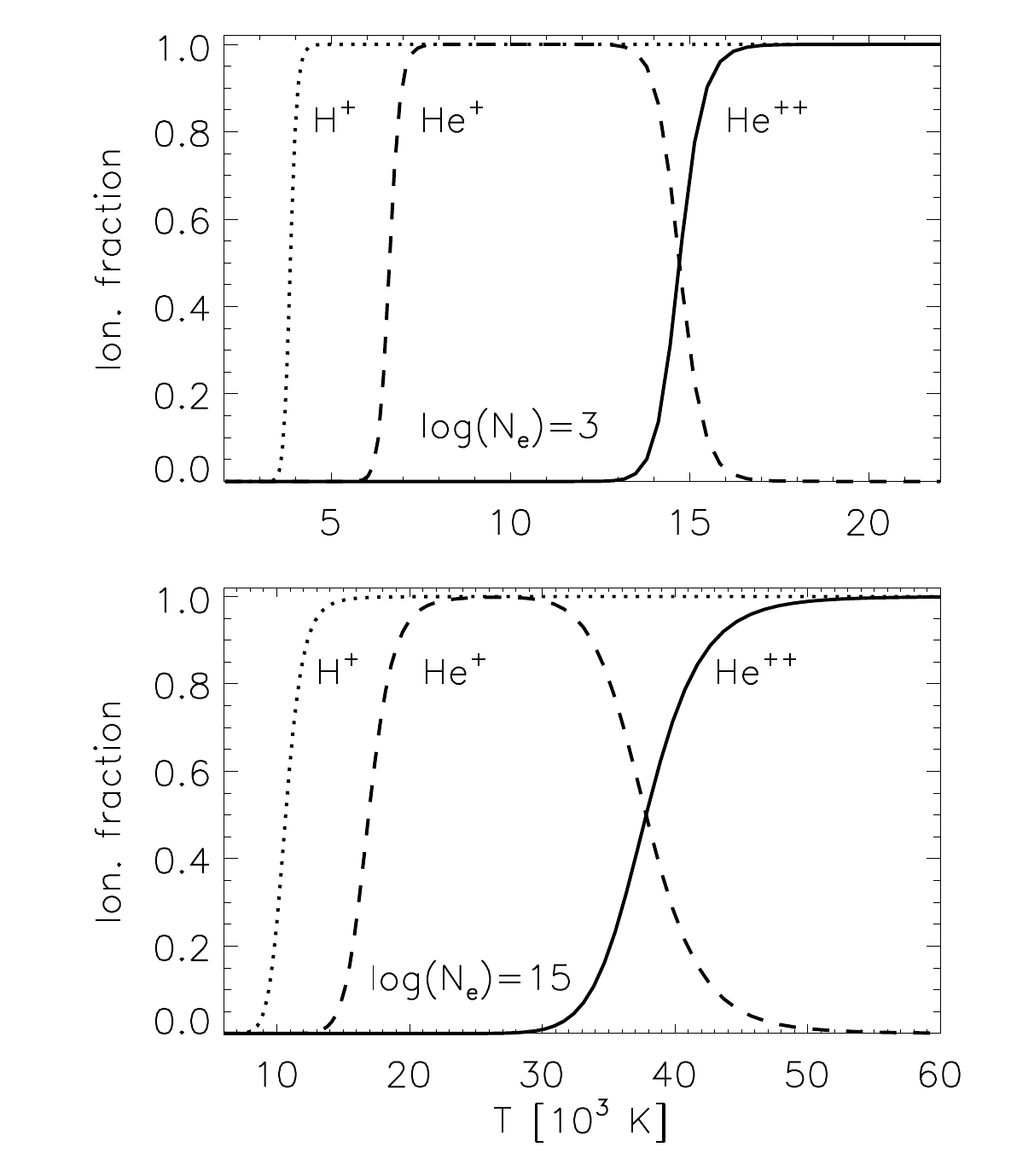}
\caption{\label{HHe_ionfrac}Hydrogen and helium ionization fractions for 
$N_e=10^3$\,cm$^{-3}$ and $10^{15}$\,cm$^{-3}$, as calculated by {\tt NEBULAR}.}
\end{figure}

\subsubsection{Other remarks}
\citet{sth95} and \citet{sts15} have provided data servers to access and interpolate 
their data tables. These servers are not directly useful for {\tt NEBULAR} because of 
their interactive nature. They were used to extract the data on the pre-defined 
$({\rm log}\,N_e,{\rm log}\,T)$ grid points. As the original grids are not strictly regular, 
{\tt Mathematica} was used to build a regular grid, enabling bicubic 2D {\tt GSL} 
interpolation for {\tt NEBULAR}. Within rounding errors, {\tt NEBULAR} yields the same 
results as the original data servers.

Recently, it has been discussed that the energy distributions of free electrons may 
not follow a purely thermal Maxwell-Boltzmann distribution \citep[e.g.][]{sts15,zzl16}. 
At this point, {\tt NEBULAR} adopts Maxwell-Boltzmann distributions throughout.

\subsection{\label{ionfracs}Ionization fractions}
The ionization fractions of H$^+$, He$^+$ and He$^{++}$ (and their number densities, $N_X$), 
are calculated for a mixed H/He plasma in collisional ionization equilibrium, solving 
the Saha equation similar to \citet{rou62}. 
Therein, the algorithm starts with a fixed matter density $\rho$, and then iteratively
obtains $N_e$; the ionization fractions follow. In the case of {\tt NEBULAR}, the 
iteration is skipped because $N_e$ is a fixed parameter. Figure \ref{HHe_ionfrac} shows 
these ionization fractions, calculated for a low and a high density mix. Alternatively,
user-defined estimates for the ionization fractions can be given to {\tt NEBULAR} 
as command-line arguments. 

\begin{figure}[t]
\includegraphics[width=1.0\hsize]{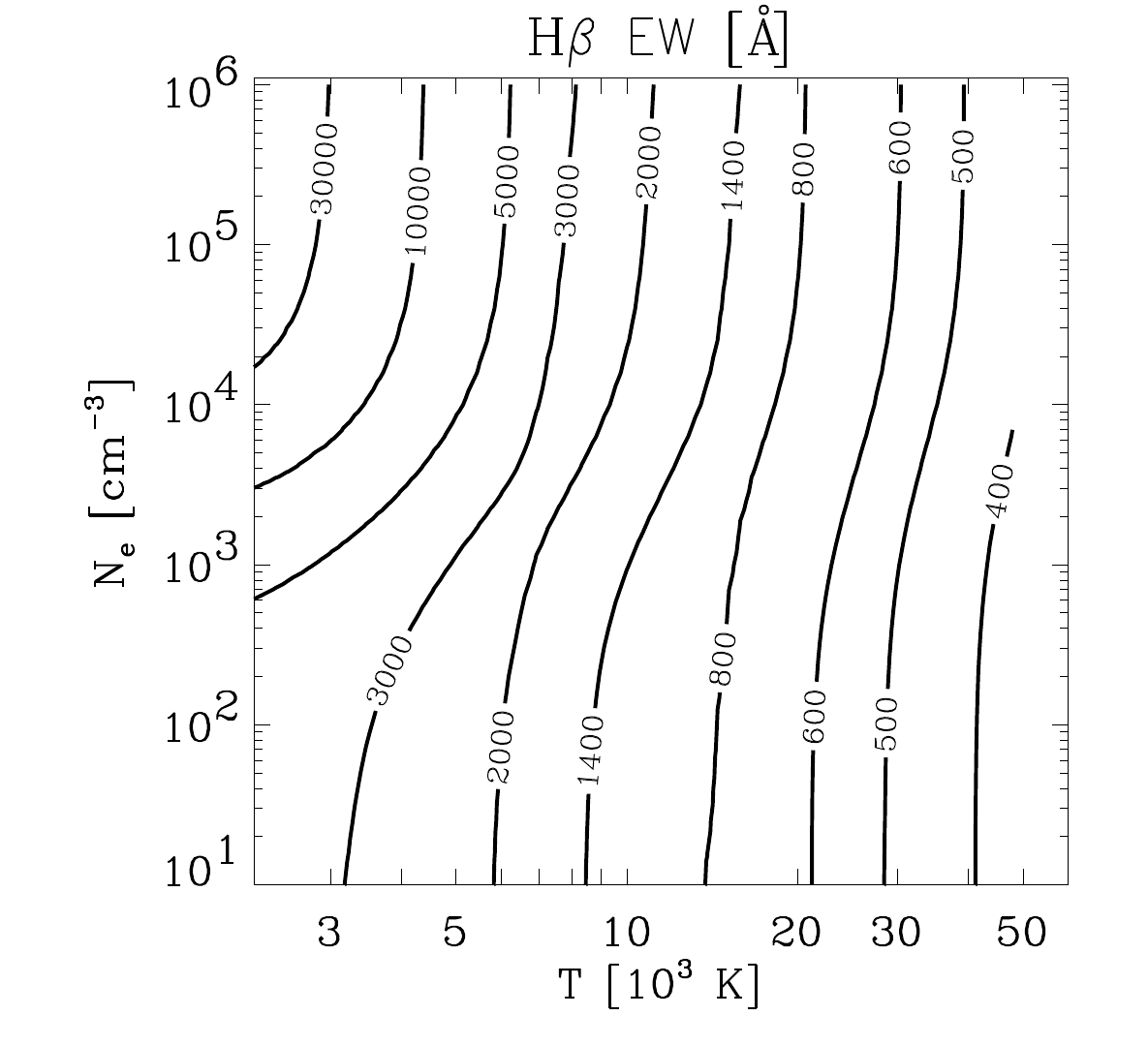}
\caption{\label{Hbeta_EW}H$\beta$ equivalent width over the full nebular continuum, as a 
  function of $N_e$ and $T$, for a helium abundance of 0.1 by parts. To create this plot, 
  {\tt NEBULAR} was run on a $(N_e,\,T)$ grid, and the EWs were evaluated for each grid point.}
\end{figure}

\section{Verification and usage}
\subsection{Consistency checks}
A series of consistency checks have been carried out to minimize the number of bugs in 
the code.

First, data tables of \citet{pen64}, \citet{dvd69}, \citet{hus87}, \citet{sth95},
\citet{ers06}, \citet{pfs12}, \citet{vhw14} and \citet{sts15} are used 
in the code. They were restructured to allow for easier and homogeneous integration in 
the software, and some of them had to be digitized manually. This is an error-prone process. 
All ($N_e$,$T$) surfaces of the derived data products were inspected in {\tt Mathematica}'s 
interactive 3D visualization module; typographical errors, and errors during restructuring 
of the multi-dimensional data tables are immediately recognizable. For example, a typographical 
mistake was found in table VI of \citet{pen64} for the \ion{He}{II} Pickering series, where it 
should read 0.39 instead of 3.39 (see Appendix \ref{corrections}). 

Second, the {\tt GSL} 2D interpolation performed on the ($N_e,\,T$) grid has been verified by 
comparing the numeric results with those given in the literature \citep[e.g.][]{ers06}. For the 
emission line spectra, \citet{sth95} and \citet{sts15} have provided dedicated data servers 
operating on the original data tables; {\tt NEBULAR}'s {\tt GSL} 2D interpolation reproduces 
numerically identical values based on the restructured tables.

Third, some figures in the reference publications compare several quantities with each other. 
For example, fig. 1 in \citet{ers06} displays the free-bound continuum emission coefficient 
for \ion{H}{I}, \ion{He}{I} and \ion{He}{II}. A similar plot is found in fig. 4.1 of 
\citet{osf06} for the combined free-bound and free-free continuum, also showing the two-photon 
spectrum. It was verified that the corresponding sub-routines in {\tt NEBULAR} reproduce these 
figures (see Appendix \ref{figcomparison} for an example).

Lastly, comparison calculations have been performed with {\tt CLOUDY} for simple
gas configurations. While {\tt CLOUDY} and {\tt NEBULAR} rely to a large extent on the
same atomic data bases, differences do exist. Most noteworthy, for example, are different
amplitudes for the helium free-bound continuum, because {\tt CLOUDY} is a photo-ionization
code, whereas {\tt NEBULAR} simply solves the Saha equation for a collision ionized gas.
To compensate for this, the ionization fractions calculated by {\tt NEBULAR} may be overridden
with command line arguments.

\subsection{Command-line arguments}
{\tt NEBULAR} is invoked on the command-line and controlled by mandatory and optional 
arguments (Table \ref{commandlineargs}). If no arguments are given, {\tt NEBULAR} reports 
its usage.

There are three mandatory arguments: a wavelength/frequency range including step size, the 
electron density $N_e$, and the electron temperature $T$. The range is assumed to be a 
wavelength range in \AA$\;$if the numeric minimum and maximum values are less than $10^7$; 
otherwise, frequencies [Hz] are assumed. Alternatively, a file containing an ASCII table 
may be provided, containing the wavelengths for e.g. each pixel of an observed spectrum.

Three example runs with {\tt NEBULAR} can be found in Appendix \ref{examples}.

\subsection{Output}
{\tt NEBULAR} produces an ASCII table with the various components and the total spectrum 
evaluated at each wavelength (or frequency). The results can be readily displayed 
in {\tt TOPCAT} \citep{tay05}. 

If a wavelength range is given, then the output spectrum will be in wavelength 
units ($j_\lambda$, erg\,s$^{-1}$\,cm$^{-3}$\,\AA$^{-1}$). If frequencies are given, 
then the spectrum will be returned in frequency units ($j_\nu$, 
erg\,s$^{-1}$\,cm$^{-3}$\,Hz$^{-1}$). Internally, all calculations are performed in frequency 
units. An excerpt of the output is shown in Table \ref{outtable}. Optionally, instead of 
$j=1/(4\pi)\,N_X\,N_e\,\gamma$, the atomic emission coefficients $\gamma$ can be returned 
\citep[e.g. to access the tabulated data by][]{ers06,sts15}. For examples, see 
Appendix \ref{examples}.

{\tt NEBULAR} also prints various parameters to the header comment section of the output file,
such as the derived ionization fractions, the ionic number densities, and the total physical 
gas density.

Pure hydrogen (helium) spectra can be produced by setting the abundance to 0 (1).

\subsection{Possible applications}
The synthesis of a full spectrum from $300-30000$\,\AA\;with 0.1\,\AA\;resolution takes a 
few seconds on a 2 GHz CPU. It is straight forward to compute e.g. the equivalent 
width of a particular emission line over the nebular continuum (Fig. \ref{Hbeta_EW}), 
and to study the relative contributions of the various continuum processes (Fig. 
\ref{hhespec} in Appendix \ref{modelspectrum}). 

Usually, the nebular continuum is weak and often undetected in extragalactic 
observations. Yet it can become significant at far-UV wavelengths because of the strong
two-photon emission. If the continuum has been fixed based on optical observations
of e.g. the H$\beta$ line, then its contribution to the \textit{GALEX} NUV and FUV bandpasses 
may be estimated \citep[the reason this software was developed,][]{sml16}. 
Another application would be the observational properties of the Balmer break at 3646\,\AA, 
as a function of spectrograph resolution and dispersion 
\citep[see Fig. \ref{balmerbreak}, and also][]{lsk09}.

\section{Summary}
{\tt NEBULAR} is a lightweight code that synthesizes the spectrum of an ideal H/He 
gas mix in (collision) ionization equilibrium over a wide range of densities, temperatures 
and wavelengths. It relies on recent atomic computations by others. It's strengths 
are mainly on the educational side, to study the response of the spectrum and its various
constituents to changes in electron temperature and density. Nonetheless, {\tt NEBULAR} 
also has scientific applications, e.g. to determine equivalent line widths and to extrapolate
fluxes in unobserved wave bands. To infer accurate physical parameters of a real, astrophysical 
nebula, full photoionization codes such as {\tt CLOUDY} must be used. 

\section*{Acknowledgments}
I thank Gary Ferland, Hai Fu and William Keel for sharing their wisdom and patiently answering 
my questions. I am indebted to the anonymous referee, for spotting mistakes in the manuscript 
and for many suggestions that substantially improved this paper.

Support for this work was provided by the National Aeronautics and Space Administration 
through \textit{Chandra} Award Number GO4-15110X (PI: M. Schirmer) issued by the 
\textit{Chandra} X-ray Observatory Center, which is operated by the Smithsonian
Astrophysical Observatory for and on behalf of the National Aeronautics Space 
Administration under contract NAS8-03060. 

\bibliography{mybib}

\begin{thebibliography}{26}
\expandafter\ifx\csname natexlab\endcsname\relax\def\natexlab#1{#1}\fi

\bibitem[{{Brocklehurst}(1971)}]{bro71}
{Brocklehurst}, M. 1971, \mnras, 153, 471

\bibitem[{{Brown} \& {Mathews}(1970)}]{brm70}
{Brown}, R.~L. \& {Mathews}, W.~G. 1970, \apj, 160, 939

\bibitem[{{Drake} {et~al.}(1969){Drake}, {Victor}, \& {Dalgarno}}]{dvd69}
{Drake}, G.~W., {Victor}, G.~A., \& {Dalgarno}, A. 1969, Physical Review, 180,
  25

\bibitem[{{Ercolano} \& {Storey}(2006)}]{ers06}
{Ercolano}, B. \& {Storey}, P.~J. 2006, \mnras, 372, 1875

\bibitem[{{Ferland} {et~al.}(2013){Ferland}, {Porter}, {van Hoof}, {Williams},
  {Abel}, {Lykins}, {Shaw}, {Henney}, \& {Stancil}}]{fph13}
{Ferland}, G.~J., {Porter}, R.~L., {van Hoof}, P.~A.~M., {et~al.} 2013, \rmxaa,
  49, 137

\bibitem[{{Golovatyj} {et~al.}(1997){Golovatyj}, {Sapar}, {Feklistova}, \&
  {Kholtygin}}]{gsf97}
{Golovatyj}, V.~V., {Sapar}, A., {Feklistova}, T., \& {Kholtygin}, A.~F. 1997,
  Astronomical and Astrophysical Transactions, 12, 85

\bibitem[{{Guzm{\'a}n} {et~al.}(2016){Guzm{\'a}n}, {Badnell}, {Williams}, {van
  Hoof}, {Chatzikos}, \& {Ferland}}]{gbw16}
{Guzm{\'a}n}, F., {Badnell}, N.~R., {Williams}, R.~J.~R., {et~al.} 2016, \mnras

\bibitem[{{Hirata}(2008)}]{hir08}
{Hirata}, C.~M. 2008, \prd, 78, 023001

\bibitem[{{Hummer} \& {Storey}(1987)}]{hus87}
{Hummer}, D.~G. \& {Storey}, P.~J. 1987, \mnras, 224, 801

\bibitem[{{Laporte} \& {Meggers}(1925)}]{lam25}
{Laporte}, O. \& {Meggers}, W.~F. 1925, Journal of the Optical Society of
  America, 11, 459

\bibitem[{{Lintott} {et~al.}(2009){Lintott}, {Schawinski}, {Keel}, {van Arkel},
  {Bennert}, {Edmondson}, {Thomas}, {Smith}, {Herbert}, {Jarvis}, {Virani},
  {Andreescu}, {Bamford}, {Land}, {Murray}, {Nichol}, {Raddick}, {Slosar},
  {Szalay}, \& {Vandenberg}}]{lsk09}
{Lintott}, C.~J., {Schawinski}, K., {Keel}, W., {et~al.} 2009, \mnras, 399, 129

\bibitem[{{Nussbaumer} \& {Schmutz}(1984)}]{nus84}
{Nussbaumer}, H. \& {Schmutz}, W. 1984, \aap, 138, 495

\bibitem[{{Osterbrock} \& {Ferland}(2006)}]{osf06}
{Osterbrock}, D.~E. \& {Ferland}, G.~J. 2006, {Astrophysics of gaseous nebulae
  and active galactic nuclei} (University Science Books)

\bibitem[{{Pengelly}(1964)}]{pen64}
{Pengelly}, R.~M. 1964, \mnras, 127, 145

\bibitem[{{Pengelly} \& {Seaton}(1964)}]{pes64}
{Pengelly}, R.~M. \& {Seaton}, M.~J. 1964, \mnras, 127, 165

\bibitem[{{Porter} {et~al.}(2012){Porter}, {Ferland}, {Storey}, \&
  {Detisch}}]{pfs12}
{Porter}, R.~L., {Ferland}, G.~J., {Storey}, P.~J., \& {Detisch}, M.~J. 2012,
  \mnras, 425, L28

\bibitem[{{Porter} {et~al.}(2013){Porter}, {Ferland}, {Storey}, \&
  {Detisch}}]{pfs13}
{Porter}, R.~L., {Ferland}, G.~J., {Storey}, P.~J., \& {Detisch}, M.~J. 2013,
  \mnras, 433, L89

\bibitem[{{Rouse}(1962)}]{rou62}
{Rouse}, C.~A. 1962, \apj, 135, 599

\bibitem[{{Schirmer} {et~al.}(2016){Schirmer}, {Malhotra}, {Levenson}, {Fu},
  {Davies}, {Keel}, {Torrey}, {Bennert}, {Pancoast}, \& {Turner}}]{sml16}
{Schirmer}, M., {Malhotra}, S., {Levenson}, N.~A., {et~al.} 2016, ArXiv
  e-prints, arXiv1607.06481

\bibitem[{{Spitzer}(1956)}]{spi56}
{Spitzer}, L. 1956, {Physics of Fully Ionized Gases} (Interscience Publishers)

\bibitem[{{Storey} \& {Hummer}(1995)}]{sth95}
{Storey}, P.~J. \& {Hummer}, D.~G. 1995, \mnras, 272, 41

\bibitem[{{Storey} \& {Sochi}(2015)}]{sts15}
{Storey}, P.~J. \& {Sochi}, T. 2015, \mnras, 446, 1864

\bibitem[{{Taylor}(2005)}]{tay05}
{Taylor}, M.~B. 2005, in Astronomical Society of the Pacific Conference Series,
  Vol. 347, Astronomical Data Analysis Software and Systems XIV, ed.
  P.~{Shopbell}, M.~{Britton}, \& R.~{Ebert}, 29

\bibitem[{{van Hoof} {et~al.}(2014){van Hoof}, {Williams}, {Volk}, {Chatzikos},
  {Ferland}, {Lykins}, {Porter}, \& {Wang}}]{vhw14}
{van Hoof}, P.~A.~M., {Williams}, R.~J.~R., {Volk}, K., {et~al.} 2014, \mnras,
  444, 420

\bibitem[{{Vrinceanu} {et~al.}(2012){Vrinceanu}, {Onofrio}, \&
  {Sadeghpour}}]{vos12}
{Vrinceanu}, D., {Onofrio}, R., \& {Sadeghpour}, H.~R. 2012, \apj, 747, 56

\bibitem[{{Zhang} {et~al.}(2016){Zhang}, {Zhang}, \& {Liu}}]{zzl16}
{Zhang}, Y., {Zhang}, B., \& {Liu}, X.-W. 2016, \apj, 817, 68

\end{thebibliography}

\appendix
\section{{\tt NEBULAR} example runs}\label{examples}
This Appendix contains some example figures and the {\tt NEBULAR} syntax used to compute
the data.
\subsection{\label{figcomparison}Extracting atomic data - emission coefficients}

\begin{figure}[h]
\includegraphics[width=1.0\hsize]{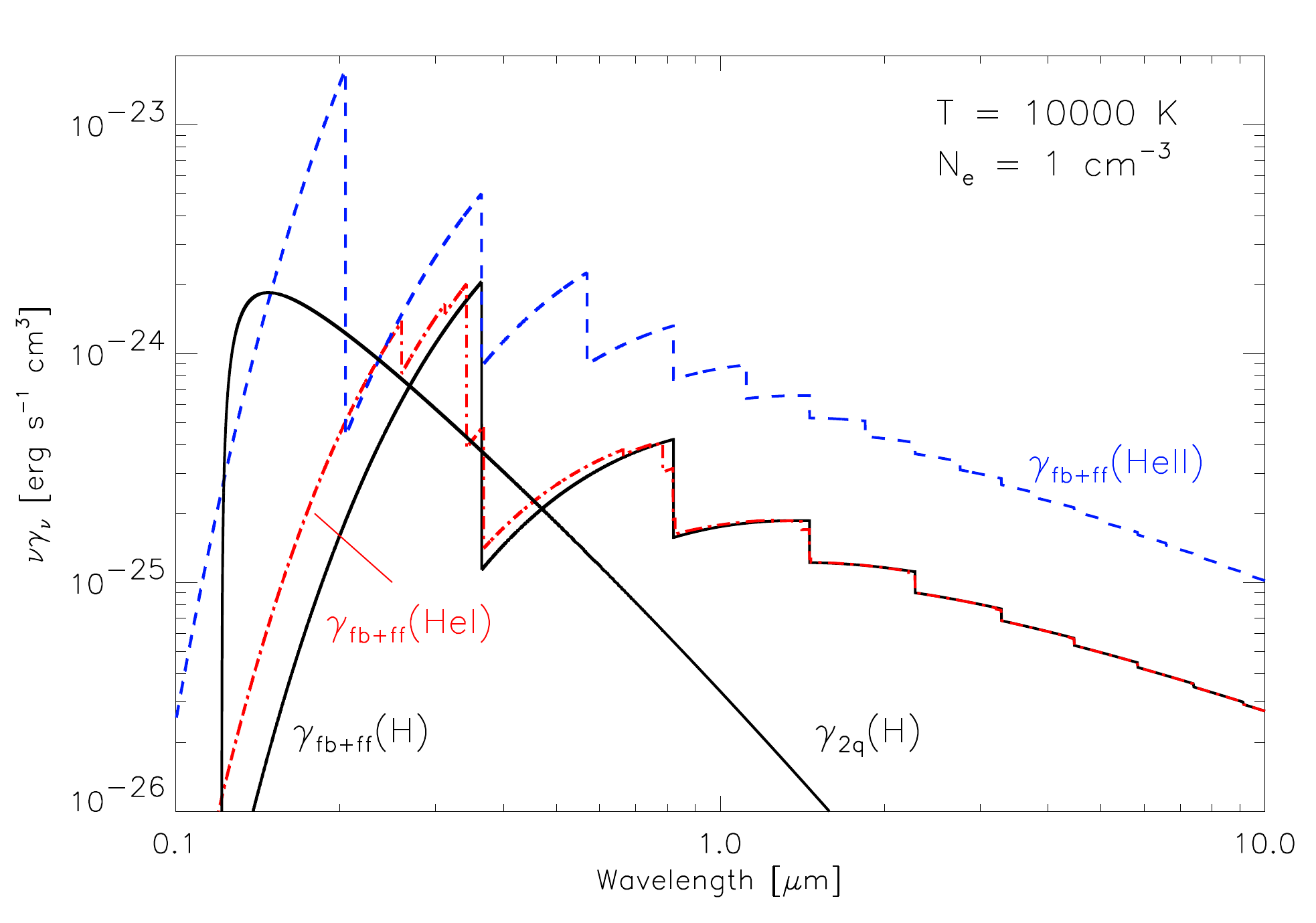}
\caption{\label{fig41}Combined free-bound and free-free continuum for \ion{H}{I} (jagged black 
  line), \ion{He}{I} (dash-dotted, red), \ion{He}{II} (dashed, blue), and the two-photon 
  continuum for \ion{H}{I} (smooth black line), to be compared with fig. 4.1 in \citet{osf06}.
  Small differences between our data and that of \citet{osf06} are attributed to the use of 
  recent data from \citet{ers06} and \citet{vhw14}.}
\end{figure}

The data for Fig. \ref{fig41} can be reproduced with the following two commands:\\\\
\indent\indent{\tt > ./nebular -r 1e13 6e15 1e11 -t 10000 -n 1 -u 2 -k -o out.dat}\\
\indent\indent{\tt > awk '(\$0!$\sim$/\#/) \{print \$2, \$4+\$10, \$5+\$11, \$6+\$12, \$7\}' out.dat > plot.dat}\\

The first command calculates the spectrum. Option {\tt -r} specifies the frequency 
range and step size, needed to get the result in frequency units. Option {\tt -u 2} 
specifies that the computations yield the emission coefficients ($\gamma_{\rm fb}$ etc.),
i.e. the atomic data tabulated in e.g. \citet{ers06}, multiplied by frequency $\nu$. 
Option {\tt -k} suppresses the calculation of emission lines. A summary of the 
syntax is given in Table \ref{commandlineargs}.

The second command line is to extract the wavelengths, the sums of free-bound and free-free 
emission for the three ionic species, and the hydrogen two-photon spectrum, respectively.

\subsection{Calculating a model spectrum}\label{modelspectrum}
\begin{figure*}[h]
\includegraphics[width=1.0\hsize]{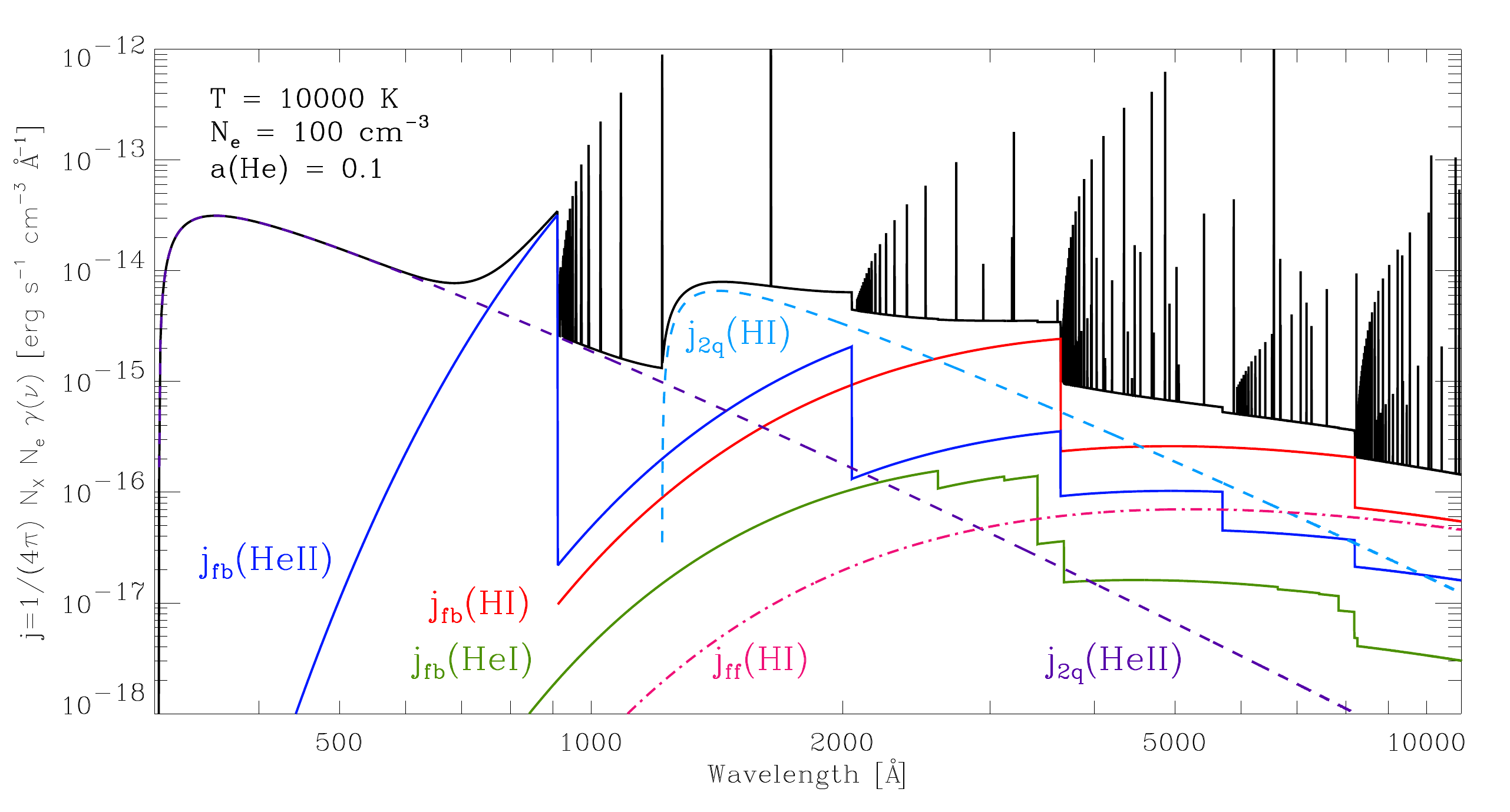}
\caption{\label{hhespec}Total emissiion coefficient $j$ for a hydrogen helium gas mix
  (top thick solid line, including the emission line series). The other three solid 
  lines with the characteristic ionization edges represent the free-bound contributions 
  by \ion{H}{I}, \ion{He}{I} and \ion{He}{II}. The two dashed lines display the two-photon 
  continua (\ion{H}{I} and \ion{He}{II}), and the lower dash-dotted line the \ion{H}{I} 
  free-free emission.}
\end{figure*}

Figure \ref{hhespec} shows a model spectrum. The difference to Fig. \ref{fig41} is that
{\tt NEBULAR} multiplied the various emission coefficients by the electronic and ionic 
densities. A wavelength range (as compared to a frequency range) is chosen as input,
and thus the result is in wavelength units (as opposed to frequency units in the previous 
example).

The data for Figure \ref{hhespec} were calculated with\\\\
\indent\indent {\tt > ./nebular -r 300 12000 1 -t 14000 -n 100}\\

If one were to compare this spectrum with a real observation, then it just has to be
multiplied with a constant factor (effectively, integrating over the nebula's 
volume and folding in the luminosity distance). This would yield the observable 
spectral flux density, $f_\lambda(\lambda)$. 

If the observed spectrum is on an arbitrary 
wavelength scale, then {\tt NEBULAR} can resample the spectrum on the same grid:\\\\
\indent\indent {\tt > ./nebular -i <ASCII wavelength table> -t 14000 -n 100}

\newpage
\subsection{Visualizing the pseudo-continuum near the Balmer break}

\begin{figure*}[h]
\includegraphics[width=1.0\hsize]{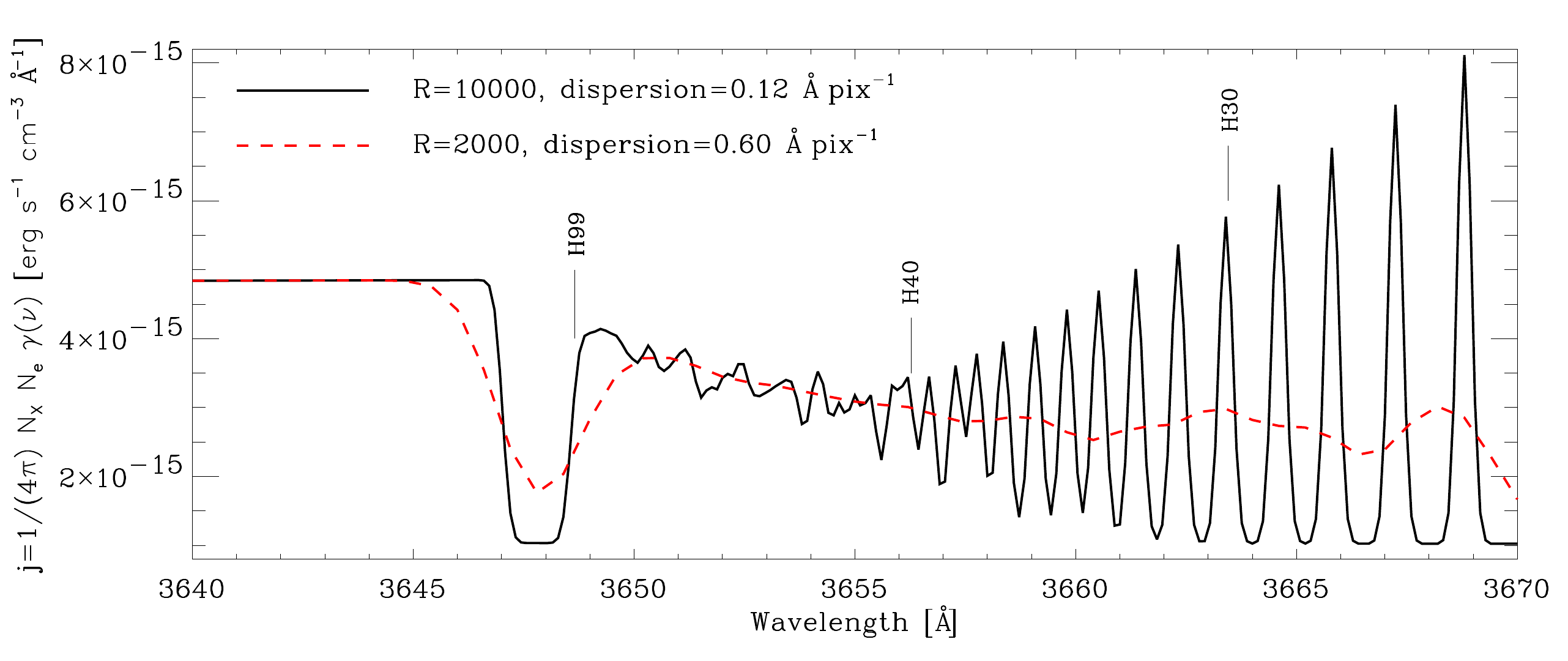}
\caption{\label{balmerbreak}The hydrogen Balmer break at $3646$\,\AA, as seen by two 
  model spectrographs with resolutions $R=2000$ and $R=10000$, respectively. The confluent 
  high-order Balmer series forms a pseudo-continuum, masking the Balmer break. The broad 
  minimum between 3647\,\AA$\;$and 3649\,\AA$\;$is artificial, because upper levels 
  higher than $n=99$ are not available in {\tt NEBULAR}.}
\end{figure*}

The data for this plot can be generated by smoothing the model spectrum with a Gaussian 
kernel, to mimic the finite spectrograph resolution, $R=\lambda/\Delta\lambda$. The FWHM
of the kernel is set to $\Delta\lambda$. The spectral dispersion is controlled by the 
wavelength step size (third numeric argument for option {\tt -r}). For the $R=10000$ 
and $R=2000$ examples, dispersions of 0.12 and 0.60\,\AA$\,{\rm pix}^{-1}$ are used:\\\\
\indent\indent {\tt > ./nebular -r 3640 3670 0.60 -t 10000 -n 100 -w 1.80 -b -o R2000.dat}\\
\indent\indent {\tt > ./nebular -r 3640 3670 0.12 -t 10000 -n 100 -w 0.36 -b -o R10000.dat}\\

Options {\tt -w} and {\tt -b} specify the kernel width and suppress the output of header 
lines, respectively.

\section{Corrections to the literature}\label{corrections}
This is a list of errors (mostly of typographical nature) found in the literature:
\begin{itemize}
\item{In Table VI of \citet{pen64} for the \ion{He}{II} Pickering series,
for $n=19$, $t'=1$ and Case B, it should read 0.39 instead of 3.39.}
\item{\citet{pes64} erroneously have a factor of $\pi^2$ in the denominator
  for the Debye radius (their eq. 33). Correct would be a single factor $\pi$.
  Subsequent derivations and numeric evaluations, however, are correctly calculated.}
\item{In their eq. (3.40), \citet{bro71} also have a factor of $\pi^2$ in the 
  denominator for the Debye radius, and carry it over in their calculations of the
  collision rates. \citet{spi56} is cited for the equation, yet in that original 
  work the dependence on $\pi$ is linear. {\tt CLOUDY} (v13.03) has the same wrong
  dependence on $\pi^2$ in the calculation of the collision rates, a bug that has been
  fixed in v15.}
\item{In Table 4.2 of \citet{osf06} the \ion{H}{I} Paschen-line series,
the values tabulated for $j_{{\rm P}\delta}/j_{{\rm H}\beta}$ actually refer to 
$j_{{\rm P}\epsilon}/j_{{\rm H}\beta}$. This table has been derived from 
\citet{pen64}, where this error is absent.}
\item{In Table 4.3 of \citet{osf06} the relative intensities with 
respect to \ion{He}{II}$\lambda4686$ are mistakenly a factor of 10 
too low for $n\rightarrow2$ and $T=5000$, 10000 and 20000\,K. This table
has been reproduced from \citet{sth95}, where this error is absent.}
\end{itemize}

\section{Supplementary tables}
\begin{table*}[h]
\caption{\label{commandlineargs}Command-line arguments for {\tt NEBULAR}.}
\begin{tabular}{lll}
\tableline\tableline
\smallskip
Args. & Value & Purpose\\
\tableline
\tableline
{\tt -r} & min, max, delta [\AA$\;$ or Hz] & wavelength / frequency range and step size;\\
         & & the output will have frequency or wavelength units\\
{\tt -i} & input wavelength/frequency table & alternative to {\tt -r}; {\tt ASCII} format\\
{\tt -t} & electron temperature [K]& \\
{\tt -n} & electron number density [cm$^{-3}$]& \\
{\tt -f} & H$^+$, He$^+$, He$^{++}$ ion. fractions & optional; calculated internally if omitted\\
{\tt -o} & output file name & optional; default: {\tt nebular\_spectrum.dat}\\
{\tt -a} & helium abundance ratio by parts & optional; default: 0.10 \\
{\tt -c} & A or B & optional; Case A/B for emission lines; default: B\\
{\tt -w} & FWHM [\AA$\;$or Hz] & optional; FWHM of a Gaussian kernel to convolve the spectrum\\
{\tt -b} & $-$ & optional; suppresses header lines if given\\
{\tt -k} & $-$ & optional; suppresses emission lines if given (faster execution)\\
{\tt -u} & 0, 1, or 2 & 0: returns $j=1/(4\pi)\,N_X\,N_e\,\gamma$ (model spectra)\\
         & & 1: returns $\gamma$ (atomic data)\\
         & & 2: returns $\nu\gamma$ \\
\tableline
\end{tabular}
\end{table*}

\begin{table*}
\caption{\label{outtable}Resulting output of {\tt NEBULAR} (abridged). The first two columns 
contain frequencies and wavelengths. The third column is the total spectrum, either $f_\nu$ or 
$f_\lambda$. The remaining columns give the free-bound, two-photon, free-free and emission-line 
components for \ion{H}{I}, respectively (data for \ion{He}{I} and \ion{He}{II} are not shown).}
\begin{tabular}{lrrrrrr}
\tableline\tableline
$\nu$ [Hz] & $\lambda$ [\AA] & $f_{\lambda}$ & $j_{\rm fb}$(\ion{H}{I}) &
 $j_{\rm 2q}$(\ion{H}{I}) & $j_{\rm ff}$(\ion{H}{I}) & $j_{\rm nn'}$(\ion{H}{I})\\
\tableline
\tableline
6.16603e+14 & 4862.0 & 6.46661e-16 & 2.92421e-16 & 2.75241e-16 & 3.66144e-17 & 0 \\
6.16578e+14 & 4862.2 & 6.46636e-16 & 2.92431e-16 & 2.75203e-16 & 3.66159e-17 & 0 \\
6.16552e+14 & 4862.4 & 6.46612e-16 & 2.92442e-16 & 2.75166e-16 & 3.66174e-17 & 0 \\
6.16527e+14 & 4862.6 & 6.46587e-16 & 2.92452e-16 & 2.75128e-16 & 3.66188e-17 & 0 \\
6.16502e+14 & 4862.8 & 4.42298e-12 & 2.92463e-16 & 2.75090e-16 & 3.66203e-17 & 4.42234e-12\\
6.16476e+14 & 4863.0 & 6.46538e-16 & 2.92474e-16 & 2.75053e-16 & 3.66218e-17 & 0 \\
6.16451e+14 & 4863.2 & 6.46514e-16 & 2.92484e-16 & 2.75015e-16 & 3.66233e-17 & 0 \\
6.16426e+14 & 4863.4 & 6.46490e-16 & 2.92495e-16 & 2.74978e-16 & 3.66247e-17 & 0 \\
6.16400e+14 & 4863.6 & 6.46465e-16 & 2.92506e-16 & 2.74940e-16 & 3.66262e-17 & 0 \\
6.16375e+14 & 4863.8 & 6.46441e-16 & 2.92516e-16 & 2.74902e-16 & 3.66277e-17 & 0 \\
\tableline
\end{tabular}
\end{table*}

\begin{table*}
\caption{Effective recombination coefficients to \ion{H}{I} $2^2$S (upper half) and 
  \ion{H}{I} $2^2$P (lower half), see also Sect. \ref{twophoton}. Extracted and reproduced 
  from table 8 in \citet{hus87}. The total recombination coefficient for the $n=2$ level 
  is obtained by adding the $2^2$S and $2^2$P values.
\label{recombcoeff_hydrogen}}
\begin{tabular}{lccccccccc}
\tableline\tableline
& \multicolumn{9}{c}{${\rm log}\,(N_e\,/\,{\rm cm}^{-3})$}\\
      & 2 & 3 & 4 & 5 & 6 & 7 & 8 & 9 & 10\\
\tableline
\smallskip
T [K] & \multicolumn{9}{c}{$\alpha^{\rm eff}_{{\rm HI}\,2^2{\rm S}} [{\rm cm}^3\,{\rm s}^{-1}]$}\\
\tableline
\tableline
1000  & 3.594e-13 & 3.701e-13 & 3.891e-13 & 4.226e-13 & 4.814e-13 & 5.887e-13 & 7.959e-13 & 1.163e-12 & 1.842e-12\\
3000  & 1.843e-13 & 1.861e-13 & 1.894e-13 & 1.950e-13 & 2.047e-13 & 2.211e-13 & 2.491e-13 & 2.846e-13 & 3.091e-13\\
5000  & 1.334e-13 & 1.341e-13 & 1.354e-13 & 1.377e-13 & 1.416e-13 & 1.481e-13 & 1.592e-13 & 1.710e-13 & 1.679e-13\\
7500  & 1.020e-13 & 1.023e-13 & 1.029e-13 & 1.039e-13 & 1.057e-13 & 1.087e-13 & 1.137e-13 & 1.182e-13 & 1.101e-13\\
10000 & 8.373e-14 & 8.390e-14 & 8.422e-14 & 8.477e-14 & 8.571e-14 & 8.729e-14 & 9.006e-14 & 9.210e-14 & 8.350e-14\\
12500 & 7.152e-14 & 7.162e-14 & 7.180e-14 & 7.213e-14 & 7.267e-14 & 7.358e-14 & 7.525e-14 & 7.621e-14 & 6.807e-14\\
15000 & 6.268e-14 & 6.274e-14 & 6.285e-14 & 6.305e-14 & 6.337e-14 & 6.392e-14 & 6.497e-14 & 6.540e-14 & 5.790e-14\\
20000 & 5.059e-14 & 5.061e-14 & 5.065e-14 & 5.072e-14 & 5.083e-14 & 5.102e-14 & 5.147e-14 & 5.144e-14 & 4.513e-14\\
30000 & 3.690e-14 & 3.690e-14 & 3.690e-14 & 3.689e-14 & 3.686e-14 & 3.682e-14 & 3.687e-14 & 3.665e-14 & 3.204e-14\\
\tableline
\smallskip
T [K] & \multicolumn{9}{c}{$\alpha^{\rm eff}_{{\rm HI}\,2^2{\rm P}} [{\rm cm}^3\,{\rm s}^{-1}]$}\\
\tableline
1000  & 1.151e-12 & 1.163e-12 & 1.187e-12 & 1.235e-12 & 1.328e-12 & 1.512e-12 & 1.912e-12 & 2.947e-12 & 5.979e-12\\
3000  & 4.854e-13 & 4.865e-13 & 4.890e-13 & 4.943e-13 & 5.049e-13 & 5.265e-13 & 5.703e-13 & 6.760e-13 & 9.467e-13\\
5000  & 3.185e-13 & 3.189e-13 & 3.197e-13 & 3.216e-13 & 3.254e-13 & 3.331e-13 & 3.489e-13 & 3.880e-13 & 4.919e-13\\
7500  & 2.251e-13 & 2.253e-13 & 2.257e-13 & 2.265e-13 & 2.281e-13 & 2.316e-13 & 2.386e-13 & 2.566e-13 & 3.080e-13\\
10000 & 1.747e-13 & 1.748e-13 & 1.750e-13 & 1.755e-13 & 1.764e-13 & 1.784e-13 & 1.823e-13 & 1.926e-13 & 2.247e-13\\
12500 & 1.429e-13 & 1.429e-13 & 1.430e-13 & 1.434e-13 & 1.440e-13 & 1.453e-13 & 1.478e-13 & 1.545e-13 & 1.771e-13\\
15000 & 1.208e-13 & 1.209e-13 & 1.210e-13 & 1.212e-13 & 1.217e-13 & 1.226e-13 & 1.243e-13 & 1.290e-13 & 1.461e-13\\
20000 & 9.220e-14 & 9.222e-14 & 9.229e-14 & 9.243e-14 & 9.272e-14 & 9.329e-14 & 9.425e-14 & 9.687e-14 & 1.080e-13\\
30000 & 6.217e-14 & 6.219e-14 & 6.223e-14 & 6.231e-14 & 6.247e-14 & 6.276e-14 & 6.319e-14 & 6.427e-14 & 7.047e-14\\
\tableline
\end{tabular}
\end{table*}

\begin{landscape}
\begin{table*}
\caption{Effective recombination coefficients to \ion{He}{II} $2^2$S (upper half) and 
  \ion{He}{II} $2^2$P (lower half). Extracted and reproduced from table 9 in \citet{hus87}.
\label{recombcoeff_helium}}
\begin{tabular}{lcccccccccccc}
\tableline\tableline
& \multicolumn{12}{c}{${\rm log}\,(N_e\,/\,{\rm cm}^{-3})$}\\
      & 2 & 3 & 4 & 5 & 6 & 7 & 8 & 9 & 10 & 11 & 12 & 13\\
\tableline
\smallskip
T [K] & \multicolumn{12}{c}{$\alpha^{\rm eff}_{{\rm HeII}\,2^2{\rm S}} [{\rm cm}^3\,{\rm s}^{-1}]$}\\
\tableline
\tableline
3000   & 8.296e-13 & 8.373e-13 & 8.552e-13 & 8.887e-13 & 9.487e-13 & 1.056e-12 & 1.250e-12 & 1.627e-12 & 2.425e-12 & 4.072e-12 & 8.155e-12 & 2.670e-11\\
5000   & 6.174e-13 & 6.207e-13 & 6.287e-13 & 6.439e-13 & 6.712e-13 & 7.189e-13 & 8.029e-13 & 9.551e-13 & 1.246e-12 & 1.737e-12 & 2.632e-12 & 5.930e-12\\
7500   & 4.859e-13 & 4.876e-13 & 4.917e-13 & 4.997e-13 & 5.140e-13 & 5.390e-13 & 5.821e-13 & 6.578e-13 & 7.949e-13 & 9.944e-13 & 1.272e-12 & 2.244e-12\\
10000  & 4.091e-13 & 4.101e-13 & 4.126e-13 & 4.176e-13 & 4.265e-13 & 4.421e-13 & 4.688e-13 & 5.149e-13 & 5.962e-13 & 7.030e-13 & 8.167e-13 & 1.246e-12\\
12500  & 3.572e-13 & 3.578e-13 & 3.596e-13 & 3.630e-13 & 3.691e-13 & 3.798e-13 & 3.980e-13 & 4.292e-13 & 4.863e-13 & 5.492e-13 & 5.996e-13 & 8.298e-13\\
15000  & 3.190e-13 & 3.195e-13 & 3.208e-13 & 3.232e-13 & 3.277e-13 & 3.355e-13 & 3.487e-13 & 3.713e-13 & 4.103e-13 & 4.542e-13 & 4.748e-13 & 6.122e-13\\
20000  & 2.659e-13 & 2.661e-13 & 2.669e-13 & 2.683e-13 & 2.710e-13 & 2.756e-13 & 2.835e-13 & 2.968e-13 & 3.197e-13 & 3.423e-13 & 3.381e-13 & 3.961e-13\\
30000  & 2.036e-13 & 2.037e-13 & 2.041e-13 & 2.047e-13 & 2.059e-13 & 2.080e-13 & 2.116e-13 & 2.176e-13 & 2.281e-13 & 2.361e-13 & 2.199e-13 & 2.313e-13\\
50000  & 1.430e-13 & 1.430e-13 & 1.431e-13 & 1.433e-13 & 1.437e-13 & 1.443e-13 & 1.454e-13 & 1.473e-13 & 1.508e-13 & 1.519e-13 & 1.352e-13 & 1.289e-13\\
100000 & 8.524e-14 & 8.524e-14 & 8.525e-14 & 8.526e-14 & 8.529e-14 & 8.531e-14 & 8.535e-14 & 8.541e-14 & 8.584e-14 & 8.512e-14 & 7.384e-14 & 6.482e-14\\
\tableline
\smallskip
T [K] & \multicolumn{12}{c}{$\alpha^{\rm eff}_{{\rm HeII}\,2^2{\rm P}} [{\rm cm}^3\,{\rm s}^{-1}]$}\\
\tableline
3000   & 2.830e-12 & 2.844e-12 & 2.866e-12 & 2.908e-12 & 2.995e-12 & 3.168e-12 & 3.516e-12 & 4.244e-12 & 5.888e-12 & 1.051e-11 & 2.671e-11 & 9.972e-11\\
5000   & 1.926e-12 & 1.931e-12 & 1.938e-12 & 1.953e-12 & 1.985e-12 & 2.047e-12 & 2.174e-12 & 2.433e-12 & 2.983e-12 & 4.390e-12 & 8.497e-12 & 2.206e-11\\
7500   & 1.407e-12 & 1.409e-12 & 1.412e-12 & 1.418e-12 & 1.432e-12 & 1.460e-12 & 1.516e-12 & 1.631e-12 & 1.869e-12 & 2.459e-12 & 4.036e-12 & 8.288e-12\\
10000  & 1.121e-12 & 1.123e-12 & 1.124e-12 & 1.127e-12 & 1.135e-12 & 1.150e-12 & 1.181e-12 & 1.246e-12 & 1.378e-12 & 1.706e-12 & 2.551e-12 & 4.567e-12\\
12500  & 9.375e-13 & 9.383e-13 & 9.392e-13 & 9.411e-13 & 9.457e-13 & 9.553e-13 & 9.753e-13 & 1.016e-12 & 1.101e-12 & 1.310e-12 & 1.846e-12 & 3.017e-12\\
15000  & 8.082e-13 & 8.088e-13 & 8.093e-13 & 8.106e-13 & 8.136e-13 & 8.202e-13 & 8.340e-13 & 8.624e-13 & 9.205e-13 & 1.067e-12 & 1.442e-12 & 2.209e-12\\
20000  & 6.366e-13 & 6.369e-13 & 6.371e-13 & 6.378e-13 & 6.394e-13 & 6.430e-13 & 6.507e-13 & 6.666e-13 & 6.989e-13 & 7.828e-13 & 1.001e-12 & 1.407e-12\\
30000  & 4.502e-13 & 4.503e-13 & 4.504e-13 & 4.507e-13 & 4.514e-13 & 4.529e-13 & 4.563e-13 & 4.633e-13 & 4.774e-13 & 5.161e-13 & 6.232e-13 & 7.972e-13\\
50000  & 2.857e-13 & 2.858e-13 & 2.858e-13 & 2.859e-13 & 2.862e-13 & 2.868e-13 & 2.880e-13 & 2.907e-13 & 2.955e-13 & 3.102e-13 & 3.570e-13 & 4.215e-13\\
100000 & 1.488e-13 & 1.488e-13 & 1.488e-13 & 1.488e-13 & 1.489e-13 & 1.491e-13 & 1.495e-13 & 1.503e-13 & 1.515e-13 & 1.552e-13 & 1.720e-13 & 1.912e-13\\
\tableline
\end{tabular}
\end{table*}
\end{landscape}

\end{document}